# An innovative, open, cloud, interoperable citizen engagement platform for smart government and users' interaction

DIEGO REFORGIATO RECUPERO[1], MARIO CASTRONOVO[2], SERGIO CONSOLI[1], TARCISIO COSTANZO[3], ALDO GANGEMI[1,4], LUIGI GRASSO[3], GIORGIA LODI[1], GIANLUCA MERENDINO[3], MISAEL MONGIOVÌ[1], VALENTINA PRESUTTI[1], SALVATORE DAVIDE RAPISARDA[2], SALVO ROSA[2], EMANUELE SPAMPINATO[5]

[1] *National Research Council (CNR), Via Gaifami 18, 95126 Catania, Italy*
[2] *Sielte, Via Cerza 4, 95027 San Gregorio di Catania, Italy*
[3] *Datanet, Syracuse, Contrada Targia 58, 96100 Syracuse, Italy*
[4] *Paris Nord University, Sorbonne Citè CNRS UMR7030, France*
[5] *Etna Hitech, Viale Africa 31, 95129 Catania, Italy*

Corresponding Author: Diego Reforgiato Recupero. Email: diego.reforgiato@istc.cnr.it

## ABSTRACT

This paper introduces an open, interoperable and cloud computing-based citizen engagement platform for the management of administrative processes of public administrations, which also increases the engagement of citizens. The citizen engagement platform is the outcome of a three-year Italian national project called PRISMA (Interoperable cloud platforms for smart-government, http://www.ponsmartcities-prisma.it/). The aim of the project is to constitute a new model of digital ecosystem that can support and enable new methods of interaction among public administrations, citizens, companies and other stakeholders surrounding cities. The platform has been defined by the media as an Italian "cloud" that allows public administrations to (i) acquire and enable any kind of application or service; (ii) enable access to open services; and (iii) enable access to a vast knowledge base represented as linked open data to be reused by a stakeholders community with the aim of developing new applications ("Cloud Apps") tailored to the specific needs of citizens. The platform has been used by Catania and Syracuse municipalities, two of the main cities of the Southern Italy, located in the Sicilian region. The fully adoption of the platform is rapidly spreading around the whole region (local developers have already used available APIs to create further services for citizens and administrations) to such an extent that other provinces of Sicily and Italy in general expressed their interest for its usage. The platform is available online and, as mentioned above, is open source and provides APIs for full exploitation.

## KEYWORDS

Smart city; smart governance; linked open data; citizen engagement; cloud computing

## 1. INTRODUCTION

Smart governance is defined as a subset of the smart city domain where an open dialogue between citizens and city officials is enabled through an ICT platform (Dameri and Rosenthal-Sabroux, 2014). Smart governance is indeed one of the smart city dimensions according to the Giffinger's model (Giffinger et al., 2007). However, the use of this term is still quite ambiguous and in this respect it must be differentiated by the approach defined in (Gil-Garcia et al., 2014), according to which smart governments implement smart governance initiatives. Besides, information gathering, dialogue enablement, decision-making and assessment are four main participatory process phases included in smart governance platforms. Moreover, giving the lack of standardization, a common smart city architecture that serves government purposes for innovation and sustainability has been defined in (Anthopoulos L., 2015).

*How can citizen engagement and open government be established in smart cities with the use of applications?*

Smart city projects and initiatives have a big impact on the quality of life of citizens. Citizens are the heart of a city and keys in several challenges cities face through on-going urbanization and demographic growth, consumption habits and increasing expectations. Therefore, they have to be at the heart of the solution. Citizens have always been insufficiently engaged, motivated and empowered to contribute. Besides, cities have not had deep knowledge of their citizens to actively engage them. With a better understanding of their motivations, cities can define effective strategies and tools to push citizens to be actors in smart city systems. The goals should be to stimulate, inform, and educate citizens so that they can act responsibly and proactively. When smartly mobilized, the effect of citizens' behaviour, choices, creativity and entrepreneurship might be enormous and have a deep impact. Therefore, by increasing innovation capabilities of the social system and by injecting advanced information technologies into it, cities become more open, innovative, efficient and manageable. Two factors play a key role in this scenario: (i) ICT, particularly as Internet becomes pervasive (not only through smart phones); (ii) the willingness to be open towards new citizen-driven initiatives that might not fit with the current administrative system.

In this context, the application of Semantic Web technologies on smart cities has an extremely high potential and impact (Bischof S. et al., 2014) and might provide a tool for the unification and facilitation of data integration from multiple heterogeneous sources. In particular, only recently, the Linked Open Data (LOD) initiative has been widely adopted and is now considered the reference practice for sharing and publishing structured data on the Web (Bizer et al., 2009). Since cities usually have large amount of heterogeneous data, Semantic Web best practices are key drivers in the definition of data reengineering, linking, formalization and use. In (Consoli S., et al., 2015A) we have presented a methodology used to collect, enrich, and publish LOD for the municipality of Catania in the context of the project PRISMA. City data have been collected and the process and issues to create a semantic data model have been analysed and reported. In more detail, the employed procedures, ontology design patterns and tools used for ensuring the semantic interoperability during the transformation process have been described. The work in (Consoli S., et al., 2015A) is included as one of the components of the smart city framework presented in this paper.

*Is there the need of a unified solution that offers several smart city applications?*

The huge amount and variety of data, applications and projects can be responsible for the creation of high entropy within the smart city domain. It often occurs that several actors within a city independently start to develop their own smart initiatives and projects using technological solutions. As an example, a public hospital may realize an on-line health web-site and database, a company may supply electric cars to its employers, a municipality might replace old buses with new ones with lower $CO_2$ emission, a PA might develop a website for chuckholes reporting, and so on. This example shows four different smart actions performed by four different entities that use ICT solutions to improve quality of life in urban spaces, reduce pollution, $CO_2$ emission and energy consumption, and to stimulate the reporting of uneven road surface preventing or decreasing chances of accidents. However, the four actions are not embedded within a unique framework whose results can list/collect/stimulate such initiatives, synergistically cooperate towards common goals, and communicate to the citizens the improved smartness of their city.

*Can smart city end-to-end cloud-based applications enhance the public administrations (PAs) savings?*

The use of cloud computing technologies can increase and improve the final outcome of smart cities solutions. International Telecommunication Union (ITU) (ITU, 2015) has collected a broad theoretical background in order to provide a smart sustainable city ICT meta-architecture that can offer several benefits: (i) lower software development, support and maintenance costs, (ii) provide more application portability and (iii) interoperability, (iv) enhance smart services and (v) shorten time-to-market for them. The cloud offers significant computational power[1] for decision-making and policy development systems and management of data coming from heterogeneous sources and from different domains. However, the cloud computing paradigm presents some technical challenges due to the interoperability of the cloud systems and to the adoption of reference standards.

In this paper we present an innovative, open and interoperable cloud computing citizen engagement platform for smart government that responds to the three questions written above. Our platform enables the construction of models and software involving urban and metropolitan dimension of PAs, and vertical scalable applications accessible according to self-service models. The platform simplifies and encourages the promotion and use of ICT technologies by citizens, industries and PAs. Moreover, it ensures interoperability in terms of possibility of interactions with other systems either cloud or not (it is a cloud-to-cloud infrastructure). To this purpose, the platform exposes a set of APIs that are shared among the entire platform providers in order to move automatically resources and data from a platform to another.

As any cloud-computing platform, it includes three layers: Infrastructure as a Service (IaaS), Platform as a Service (PaaS) and Software as a Service (SaaS). Whereas the IaaS layer is the backbone concerning the cloud architecture, PaaS and SaaS layers deal directly with the knowledge that can be exploited to improve productivity and to create new products, services, systems, and process. The development of the platform has been conducted according to the principles of the open source paradigm and each component has been developed using open source solutions that this paper will detail.

We provide information of each layer in the following sections. The produced platform is an outcome of the PRISMA project, a three-year Italian research project whose goal is to develop an "Italian cloud" for the needs of the PA that can be sustainable over time. The project aims at supporting the definition and adoption of new business models that allow PAs to (i) acquire and port immediately to the cloud every application and service, and (ii) let LOD be exploited by small and medium enterprises in order to create new applications capable of delivering services tailored to the specific needs of the citizens.

This paper is organized as follows. Section 2 includes background and related work. Section 3 presents the case study we propose in this paper. It includes the general architecture of the platform developed in PRISMA, the IaaS, PaaS and SaaS layers, the four applications developed within the SaaS layer and the quality, importance and impact of the presented platform with its applications and services. Finally, Section 4 ends the paper with discussions, limitations of the platform and future directions where we are headed.

## 2. BACKGROUND

Population has been moving from rural to urban places and this shift is projected to continue for the next couple of decades[2]. As a consequence, more than half of World's population now lives in urban areas (Dirks et al., 2010) (Dirks et al., 2009). Today, in fact, 78% of European citizens live in cities and 85% of the EU's gross domestic product is generated in cities. This concentration of people inevitably tends to create new kind of problems that cities need to solve. Ensuring acceptable living conditions within the context of such a rapid urban population growth triggers many cities to find smarter ways to manage the new challenges that might arise. Besides, in the international context, the concept of smart city has emerged and has been adopted by several institutions (European Commission, OECD, UNDP, Setis-EU, ANSI, etc.) in order to achieve part of the objectives established in the Kyoto Protocol.

---

[1] For reasons why smart cities need cloud services and the importance of computational power have a read at http://www.ubmfuturecities.com/author.asp?section_id=234&doc_id=526607
[2] http://www.unfpa.or

Smart cities (Chourabi, et al., 2012) use a collection of smart computing technologies to enhance quality and performance of urban services, to reduce costs and resource consumption, and to engage more effectively and actively with their citizens. Smart computing can be thought of as a new generation of integrated networks, software, and hardware technologies that provide IT systems with advanced analytics to help users to adopt better decisions that will optimize business processes. Information and communication technologies are key drivers within the smart city initiatives (Hollands, 2008). Their adoption might provide opportunities for actors around the city and, at the same time, enhance the management of a city. An important issue that obstacles the transformation from an ordinary city to a smart city is represented by the interaction with political and institutional components (Mauher and Smokvina, 2006) such as city council, city government, etc., and external pressures (political agendas and politics) that might affect the process and results of IT initiatives. The policy context is crucial for the adoption and use of information systems to the widest degree possible. Sometimes policies have to be changed by innovative governments because innovation cannot happen without a normative that can enable it. The shared goal of smart cities is to increase their common problem-solving capabilities for the benefit of citizens and PAs. Clearly, the smart city paradigm has strong implications in the PA management, in the way of doing politics, and carrying out the interactions and relations among politicians, citizens and public servants. Open Government's principles (Geiger and Lucke, 2011) such as transparency, participation, and collaboration are of utmost importance for the integration of the members of a city within the smart city paradigm.

The development of smart cities has become a major issue over the past decade. The introduction of IBM's Smart Planet and CISCO's Smart Cities and Communities programs had a strong impact on the government economies and their potentials have captured public's attention. Since then, the smart city paradigm has been seen as a fundamental component of the global knowledge economy. The understanding of the smart cities domain, focusing on the governance, modeling and analysis of the transition and contribution towards the sustainability of urban development has been explored in (Deakin, 2013).

Linking and upgrading infrastructures, technologies and services in important urban sectors such as transport, buildings, energy, ICT, in a smart way will improve quality of life, competitiveness and sustainability of cities. This is a strong growth market, estimated globally to be worth €1.3 trillion in 2020[3]. Markets are often fragmented, missing out on their full economic potential. Many innovative solutions require new business models and financing solutions to keep low potential risks. Since demand for better infrastructures and services is high and still increasing but public budget is under pressure, knowledge needs to be shared effectively and capacities to be developed. Several cities have already started to invest resources and to participate in projects with the aim of serving citizens and improving their quality of life (Giffinger et al., 2007) (Odendaal, 2003).

A smart city model would include an aggregation of technologies and processes (Deakin, 2013) such as Big Data management, the Internet of Things, sensor networks, smart devices, embedded systems, cloud computing technologies, among others. They have been having a strong impact on the evolution of medicine, transport, environment, business and government by introducing new type of knowledge processes such as information collection and processing, real-time forecasting and alerting, collective and crowdsourced intelligence, cooperative distributed problem-solving and learning (Velosa and Tratz-Ryan, 2015). An emergent direction is provided by ambient-assisted living (AAL) technologies that involve challenges (and plenty of data) for the adult healthcare system.

---

[3] http://tech.firstpost.com/news-analysis/internet-of-things-network-to-be-worth-3-trillion-by-2020-with-30-billion-connected-devices-says-idc-240972.html

On the other hand, ICT today represents a pervasive solution for urban environments (Anthopoulous and Vakali, 2012) providing the necessary features for the sustainability and resilience of a smart city. ICT has been considered as the main enabler that can transform data into specific knowledge and useful information. Among the features that ICT provides for such utilization there are smart phones, sensor nets and smart household appliances. The adoption of even smarter hardware can enable the so-called Internet of Things (IoT) and thus provide increasing amount of data for the user and the environment. However, as mentioned by authors in (Zanella et al., 2014), building a general architecture for the IoT is difficult, due to the large variety of devices, link layer technologies, and services that may be involved. Urban IoT systems can have a lower degree of complexity as they are characterized by specific application domain.

Besides, a smart city comprises several and different domains (transport, energy, land use, government); therefore, the amount of data to be handled becomes huge and heterogeneous[4]. One ICT solution for such an issue is represented by Big Data technologies that can collect, process, and integrate huge amount of data in nearly real-time and share them through interoperable services deployed within a cloud environment. As an example, authors in (Khan et al., 2015) proposed a cloud-based analytics service (implemented using Hadoop[5] and Spark[6]) and a theoretical and experimental perspective on the smart cities focused on Big Data management and analysis. Their service analyzed the Bristol Open Data (in particular data about quality of life that is crime, safety, economy and employment) by identifying correlations between selected urban environment indicators and assessed positive and negative trends.

Furthermore, authors in (Batty, 2013) provided a definition of Big Data with respect to its size related to the urban domain, and described how the growth of Big Data is shifting the emphasis from longer term strategic planning to short-term. One example is the 6-month data collected during 2011-2012 about travelers using the smart Oyster card for paying traveling (buses, tube trains and over ground heavy rail) in Greater London. The analysis of such data allowed coming up with predictions, optimization of the system, and general assessment of travelers, tips and suggestions on how to improve the system on the one hand and to identify travel patterns on the other hand.

Data in the cities is crucial for smart services. There are data related to the financial status of the city itself, budgets, incomes, expenditure; data related to the transport systems such as bus, metro, rail, trams lines and timetables; data from sensors that measure traffic, humidity, noise levels, temperature, pollutants; data about the available services, geographical information, crime statistics, etc. Besides citizens today are demanding more accountability from their governments. They want to stop squander and delays in bureaucracies' processes. They would like to access online services from their homes through the web or with their mobile phones. On the one hand, with the adoption of the open data paradigm they finally can. On the other hand, open data enables software developers to transform that data into useful applications that make city services available anytime, anywhere. Authors in (Bischof et al., 2014) included some examples of how Semantic Web can be employed to give a meaningful structure to open data dealing with cities. In particular, issues around smart city data and preliminary thoughts for creating a semantic description model to describe and help discover, index and query smart city data was shown. The Open Data City Census[7] by the Knowledge foundation is a good example showing the kind of data and the related quality in the cities. One component included in the proposed platform deals with open data and Semantic Web best practices to transform and publish open data in LOD (Consoli S., et al., 2014A). In more detail, authors in (Consoli S., et al., 2014B) presented a methodology used to collect, enrich, and publish LOD for the municipality of Catania. The collected city data, their production process, issues faced in creating a semantic data model and tools used for ensuring semantic interoperability during the transformation process were presented along with discussions/suggestions on how to use the produced open data model by local stakeholders (developers and main actors).

---

[4] ITU has defined some standards related to the definition of a Smart Sustainable City. http://www.itu.int/en/ITU-T/focusgroups/ssc/Pages/default.aspx
[5] https://hadoop.apache.org/
[6] http://spark.apache.org/
[7] ES Open Data City Census http://es-city.census.okfn.org/

Urban collaborative data collection is another recent trend in smart cities. Perhaps the most known application is represented by FixMyStreet[8] (Baykurt, 2012), which is a web based application that helps citizens inform their local authorities about problems that require their attention, such as broken streetlamps, potholes, and similar problems with streets and roads in the United Kingdom, and see what reports have already been made. One more example is represented by Improve My City[9], a freely available open source platform for local governments seeking collaboration with their citizens able to manage local issues, from reporting, to administration and analysis. Differently from the platform that we present in this paper, the Improve My City platform does not exploit any cloud architecture and it is not totally free for usage but includes additional costs, which depend on the adopted plan.

Having a smart city nowadays provides opportunities for local communities, stakeholders and developers to exploit open data and create their own applications and services. The usefulness and necessity of making cities smarter is clearly acknowledged. However, it seems that business around the smart city concept is encountering obstacles to take off. Authors in (Vilajosana et al., 2013) looked into the underlying reasons for these obstacles and propose a model for smart city development based on Big Data exploitation through the API store concept. Two major observations can be drawn: (i) smart city departments need to be identified and made independent in order to isolate the political component of the improved city servicing from the underlying technologies, (ii) a three-phase smart city action plan is needed, where in phase 1 utility and revenues are generated, in phase 2 only utility-service is supported and in phase 3 a fun-leisure dimension is allowed.

## 3. THE PROPOSED CASE STUDY

The PRISMA project was started to address a smart cities and communities call of the Italian Ministry of Education, Universities and Research (MIUR)[10]. The goal of the call was to promote within South Italian regions research products dedicated to the development of smart cities. Industries, SMEs, universities and public research institutions were asked to integrate their expertise to develop highly innovative solutions that, through the most advanced technological tools, might contribute to the development of the territories, and respond to the real needs of the community to improve the quality of life of citizens.

Some of the partners of PRISMA had business relationships with Sicilian PAs, in particular the municipality of Catania that was really interested to develop a smart city platform for citizen engagement. That was in line with the strategy that the city of Catania had fixed in its previous planning to achieve some of its goals.

The architecture has been developed keeping in mind some requirements established in the project call. Two of them were in compliance with the open source paradigm and the adoption of cloud computing technologies.

In more detail, the design of the architecture of the SaaS and PaaS layers followed a top-down approach. First we looked at the requirements and then we decided which module we needed to include keeping also in mind that the platform had to follow open source principles. The interaction between the modules has been designed according to service-oriented architectures (SOA). The ESB component we decided to include performs greatly with SOA architectures. It is the module that orchestrates all the connections with the other modules. For the IaaS layer we decided to use OpenStack (Folson version first, and Grizzly version then) after analyzing a number of available platforms. The choice of using a certain number of OpenStack services was based on a set of requirements such as open source, integration with enterprise world, use of international standards, simplified release of instances or virtual machines in self-service mode, possibility to manage accounts, resource monitoring and security, and privacy guarantee. Last but not least, a community of developers from several industries manages OpenStack and, therefore, vendor lock-in issues do not exist.

---

[8] http://fixmystreet.org/
[9] http://www.improve-my-city.com/
[10] The link for such a call is http://www.ponrec.it/programma/interventi/smartcities/ (in Italian)

### 3.1 Architecture of the Platform

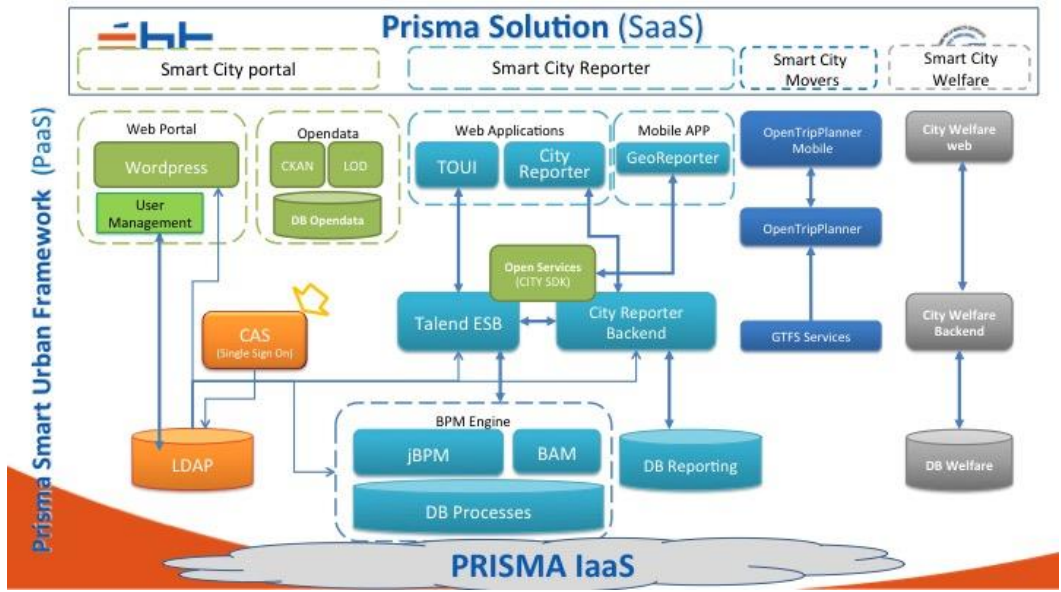

**Figure 1 - Architecture of the proposed platform**

Figure 1 depicts the architecture of the proposed platform. It consists of three main and independent parts:

- the **IaaS layer**: it is the lowest layer that includes cloud computing services and allows the user to realize an infrastructure based on servers, networks and storage in total autonomy;
- the **PaaS layer**: also known as Prisma Smart Urban Framework, it is the layer more tied to developers and their development environments since it offers software components to be used as services;
- the **SaaS layer**: also known as Prisma Solution, it consists of a collection of applications on a cloud-environment based on the IaaS and PaaS layers. It allows the end user to use the software product as a cloud component, therefore with no need of installations on local infrastructures.

This scheme allowed us to produce models and innovative implementations for processes related to PAs through the realization of scalable vertical applications, accessible according to self-service models. The applied case studies are related to the municipalities of Catania and Syracuse; however, the platform is general enough to be used for any other cities.

### 3.2 PRISMA IaaS layer

The IaaS layer is the result of a thorough activity of comparative evaluation of different IaaS components, both open source and commercials.

This layer represents the cloud overlay underneath the proposed platform for the on demand provision of virtual environment for computation and storage. The IaaS platform complies with international standards and is released under an open source license that enables its free usage. With the choice of adopting international standards, our platform stands on the line of the recommendations of the British and German governments and the American NIST, for the sake of interoperability, which are crucial to avoid vendor lock-in and promote the development of a competitive market in the field. The open license facilitates the adoption of the proposed platform by data center of PAs, industries and ICT service providers. The open source approach also facilitates transparency, security and data privacy, in accordance with the Italian laws and those of the European Union. OpenStack is the chosen open source platform on the top of which we have based our IaaS architecture.

In particular, PRISMA IaaS is a set of software components for the management of a cloud computing platform for public and private clouds. The strengths include the speed of provision of virtual machines (VM) and scalability. OpenStack cloud also allows the system to provide end users with a remote environment where the software runs as a service and permits higher reliability and scalability. This technology extends the concept of virtualization to the use of an infrastructure as a service simplifying the provision of resources to the end users.

One important best practice that the authors would like to share is the following. To ensure stability, it is highly recommendable to use OpenStack versions with software packets ready and tested for a developing/production environment. The reason is that OpenStack is made by different modules that if not well managed can create several problems.

#### 3.2.1 Technological Components of the PRISMA IaaS layer

A set of software components form the PRISMA IaaS layer. These are summarized in the following.

**Nova**[11] (OpenStack Compute) is the most important component that controls the entire platform. It is used to manage virtual machine instances and internal communication systems. Its tasks are carried out by different specialized services that work together. Nova Schedule is one service that allocates instances on physical machines (it acts as a distributed resource scheduler) whereas Nova Compute is another service that communicates with the KVM hypervisor (virtualization layer in Kernel-based Virtual Machine) installed on the physical machine to manage the various stages of an instance. **Swift**[12] (Object Storage) is a distributed storage system designed for high reliability and scalability and optimized for durability, availability and concurrency. **Glance**[13] (Image Service) is the service for the management of virtual images. Glance has a RESTful API that allows querying of VM image metadata as well as retrieval of actual image. **Keystone**[14] (Identity) provides authentication and authorization capabilities for OpenStack modules. **Horizon**[15] (Dashboard) is the access web interface to OpenStack services. **Neutron**[16] (Network) is the module for the network communication. It is an OpenStack project that provides the network as a service between different devices. Neutron performs IP (public and private) management. **Cinder**[17] (Block Storage) manages block volumes for data storage.

Figure 2 shows the interconnection of all the components just described.

---

[11] https://wiki.openstack.org/wiki/Nova
[12] https://wiki.openstack.org/wiki/Swift
[13] https://wiki.openstack.org/wiki/Glance
[14] https://wiki.openstack.org/wiki/Keystone
[15] https://wiki.openstack.org/wiki/Horizon
[16] https://wiki.openstack.org/wiki/Neutron
[17] https://wiki.openstack.org/wiki/Cinder

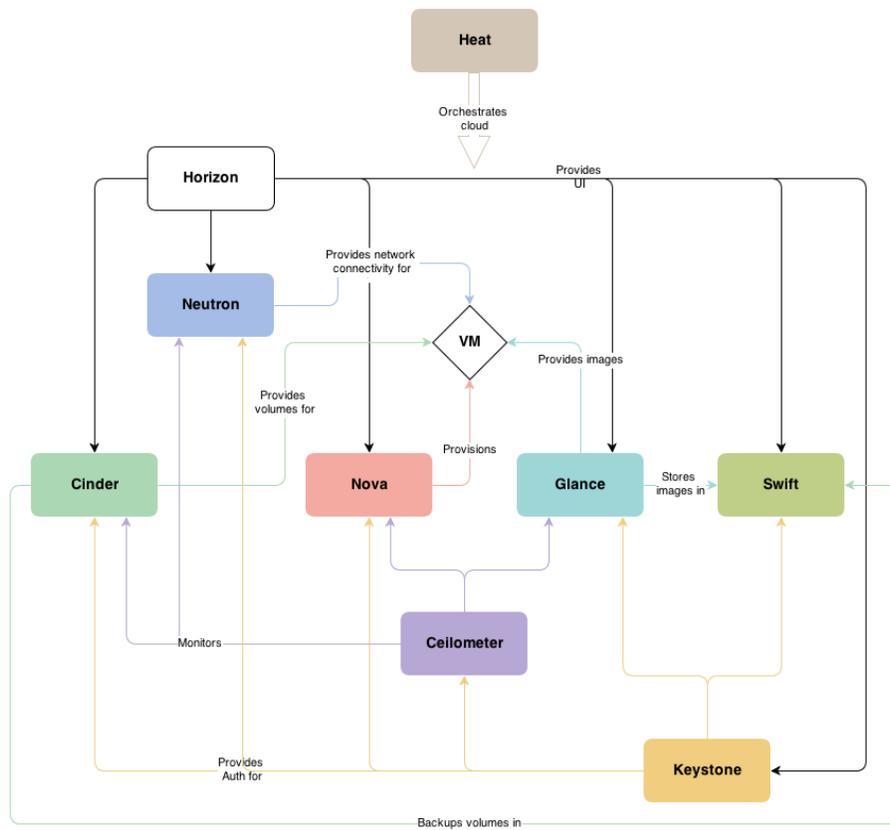

**Figure 1 – Interconnection of the components constituting the IaaS layer of the OpenStack architecture**

### 3.3 Prisma Smart Urban Framework (PaaS Layer)

Current technologies and platforms available for cloud computing are oriented to offer general virtualization services such as storage, virtual hosts, web space, etc. The idea of the Smart Urban Framework (PaaS) is instead to create a cloud specialized and oriented to e-government services. This design choice impacts the type of software architecture, which must take into account the mean features of the specific domain and the central role of the citizen, i.e. the main user of such a system, who should be the principal actor in the processes of services definition and services usage mode. At design time we planned to include features such as the kind of user, her/his degree of satisfaction, and the assessment of the services reliability. This design strategy may therefore find applicability in other contexts where the role of the user assumes the same centrality.

The PaaS layer consists of several modules, as it can be seen in Figure 1. In the following we give a high level description of the PaaS[18]. Web Portal represents the smart city portal of the PA, and the application employed for users management. TOUI executes the tasks of the Business Process Modelling (BPM) procedures whereas City Reporter (it uses City SDK standard) handles the insertions and consultations of urban reports. They are all connected with LDAP that allows users to authenticate by single sign-on method through the Central Authentication Service (CAS). The The BPM Engine provides the management of the processes, along with their design and storage.

---

[18] The reader is invited to check http://www.catania-smartcity.it/il-progetto/prisma-smart-urban-framework-paas/ for a high level description of each module.

The core of the architecture is Talend ESB that orchestrates the services among the different components.

In the following we describe in more detail each of these modules, their usage and related interconnections. We highlight (i) the benefits of each module, (ii) the modules that independently provide API access for potential developers and (iii) how the innovative solutions described in this paper can handle and manage knowledge related to PAs on the top of which is possible developing additional services and applications which affect the dynamics of the economy of a city.

### 3.3.1 BPM Engine

The BPM Engine component includes the jBPM, BAM and DB Processed modules. BPM stands for Business Process Management, a paradigm that provides a bridge between ITs, businesses and their operational processes. The three components that our BPM system employs are:

- *the process engine*, a robust platform for modeling and executing process-based applications, including business rules;
- *the business analytics*, which enables managers to identify business issues, trends, and opportunities (through reports and dashboards) and to react accordingly;
- *the content management*, which provides a system for storing and securing electronic documents, images, and other files.

Our BPM Engine aims at providing the platform with the services for the management of business processes, that is, all the urban reporting that users send through the web application and that are received and handled by city operators. The engine permits to implement software solutions that use the BPM paradigm. The BPM Engine is based on jBPM[19], an open source workflow engine written in Java that can execute business processes described in BPMN 2.0. The main motivation behind this choice is that, differently from traditional BPM engines that have a focus limited to non-technical people only, jBPM offers process management features and tools (design, implementation, simulation, execution of processes, analysis) in a way that both business users and developers can appreciate. One more component included in the BPM Engine is represented by the BAM (Business Activity Monitoring) module that monitors business activities. BAM has been implemented using Dashbuilder[20], a web application that enables creating business dashboards whose data can be extracted from heterogeneous sources of information such as JDBC databases or even text files. A MySQL database (i.e. the DB Processes module) is employed within the BPM module to store information about processes in current execution or already executed. jBPM directly communicates only with the LDAP module (see below) to access to users' credentials and Talend ESB that is the interface to BPM for the external environment: if a process wants to use BPM's services, it invokes Talend ESB that calls one or more methods of jBPM and returns its response to the caller. Note that each solution adopted for the implementation of the BPM module is open source.

### 3.3.2 LDAP

The LDAP (Lightweight Directory Access Protocol) module includes users information such as their credentials (username and passwords), groups a user belongs to and that describe which permissions a user has in each application. The module is based on the LDAP protocol, required to access and maintain distributed directory information services over an IP network. Among the different goals that the LDAP protocol accomplishes, we use it to provide a centralized server that contains users' information. The LDAP module has been developed using OpenLDAP[21] the most common, reliable and stable open source implementation of the LDAP protocol.

---

[19] http://www.jbpm.org/
[20] http://dashbuilder.org

The module is connected to WordPress, jBPM, City Reported Backend, City Welfare Backend, Talend ESB. All these modules take users credentials information from LDAP. Besides, the User Management application (a front-end of the LDAP module) inserts users and groups in LDAP and the CAS module (see below) uses LDAP as repository.

phpLDAPadmin[22] is the web-based LDAP client that provides administrative functionalities for the LDAP server.

### 3.3.3 CAS

The CAS (Central Authentication Service) module is responsible for the single sign-on that is the access control mechanism employed in multiple independent software systems. Its purpose is to permit a user to access multiple applications while providing their credentials (such as username and password) only. The CAS module uses as a backend for users' credentials the LDAP previously described. The CAS has been developed on top of Jasig[23], a very common and stable open source tool. Each web application of the architecture in Figure 1 (Wordpress, TOUI, Smart City Reporter, Smart City Welfare) uses CAS for the authentication.

### 3.3.4 Talend ESB

Talend ESB (Enterprise Service Bus) represents the integration middleware. It handles the communication between the different tools, managing all the issues such as the transformation and merging of the messages coming from different sources. It can be considered as the business layer between the modules that are interested in the business process management and the jBPM functionalities. The module has been developed using talend[24], a stable and open source solution built on top of Apache[25] ESB technologies. The module implements SOAP[26] and REST[27] web services and links the jBPM module with the TOUI and City Reporter Backend modules.

### 3.3.5 City Reporter Backend and DB reporting

The City reporter backend component enables the insertion and querying of urban reporting data through REST web services. It is integrated with LDAP for the management of the users and their interactions. JAVA is the language used to develop such a module. In particular, the Spring[28] framework has been employed: Spring Security[29], to handle the security component, and Spring MVC[30], to develop the REST services. Hibernate[31] has been used to map the Java model to the database, and therefore to allow the access to the data. MySQL is the employed Database Management System (DBMS). It represents the database of the City Reporter Backend (DB reporting). Please note that the chosen framework represents the state of art in terms of applications development in JAVA. The City Reporter Backend module is directly linked to the City Reporter Front-end, the Talend-ESB, LDAP and the DB reporting.

---

[21] http://www.openldap.org/
[22] http://phpldapadmin.sourceforge.net/wiki/index.php/Main_Page
[23] http://jasig.github.io/cas/4.1.x/index.html
[24] https://www.talend.com/resource/open-source-esb.html
[25] http://www.apache.org/
[26] https://en.wikipedia.org/wiki/SOAP
[27] https://en.wikipedia.org/wiki/Representational_state_transfer
[28] http://spring.io/
[29] http://projects.spring.io/spring-security/
[30] http://docs.spring.io/spring-framework/docs/current/spring-framework-reference/html/mvc.html
[31] http://hibernate.org/

### 3.3.6 Opendata module

This module contains the technological infrastructure used for the production and storage of open data. The module consists of a set of components. One component is CKAN (Comprehensive Knowledge Archive Network), an open source content management system (CMS) for the storage, publication and distribution of open data. The kind of data to store into the open data module depends on the SaaS layer where an instance of the Smart City Portal is installed. A further component is the LOD, which includes the methodology used to collect, enrich, and publish Linked Open Data. LOD offers the possibility of using data across different domains or organizations for purposes like statistics, analysis, maps and publications. By linking this knowledge, interrelations and correlations can be quickly understood, and new conclusions arise. Through ``Uniform Resource Identifiers'' (URIs) and the ``Resource Description Framework'' (RDF), slices of information and data can be arranged, shared, exported, and connected, and APIs, applications, and tools can be created. LOD are currently bootstrapping the Web of Data by converting existing datasets into RDF and making them available to the general public under open licenses.

We have widely described this component and methodology in (Consoli S., et al., 2015A) (Consoli S., et al., 2014A) (Consoli S., et al., 2015B) (Consoli S., et al., 2014B) (Consoli S., et al., 2014C) for the municipality of Catania. We have described there the process and issues to create a semantic model, the employed procedures, ontology design patterns and tools used for ensuring semantic interoperability during the transformation process. The data model integrates several data sources including geo-referenced data, public transportation, urban fault reporting, road maintenance and municipal waste collection. Finally, the DB open data component represents the repository we have used to host all the produced RDF/OWL semantic data. In particular, we installed and used Virtuoso[32], an open source product that provides SQL, XML and RDF/OWL data management. Virtuoso allows programmers to access the produced data and ontology that are accessible by SPARQL queries[33]. Data are stored in the RDF graph <prisma>, while the ontology is stored in the RDF graph <prisma-ont>. For programmers, the SPARQL endpoint is also accessible as a REST web service. It requires as input a user-defined SPARQL query and produces as output the query result in one of the following formats: *text/html, text/rdf +n3, application/xml, application/json, or application/rdf+xml*.

The overall LOD component is publicly available online at http://wit.istc.cnr.it/prisma/webcontent/home.html. Here the reader can browse the designed ontology by user-oriented visualizations, such as Live OWL Documentation Environment (LODE) as human-readable HTML, or WebVOWL as a force-directed graph layout. WebVOWL implements the Visual Notation for OWL Ontologies (VOWL) by providing graphical depictions for elements of the OWL Web Ontology Language that are combined to a force-directed graph layout representing the ontology. Both tools enable user-oriented visualizations and provide the description of elements of the ontology. Interaction techniques allow the user to explore the ontology and to customize the visualization.

Readers may also play with data through the dedicated SPARQL endpoint, or by means of content negotiation via Web REST services. Integration with two other data visualization tools, LodView and LodLive, is also provided. LodView provides HTML based representation of our RDF resources, able to offer a W3C standard compliant URI dereferentiation, an intuitive interface, and additional interesting features, e.g. possibility to download the selected resource in different formats (*xml, ntriples, turtle, ld+json*). LodLive is a navigator of RDF resources based on a graph layout. It is used for connecting RDF browser capabilities with the effectiveness of data graph representation.

This module also provides two further high-level visualization tools for the produced data. The first one, called Semantic Geo-Visualizer, shows geo-referenced objects on a Google map. All classes and objects (from our SPARQL endpoint) that are associated to a shape can be shown in the visualization tool. Users can select a set of geo-referenced objects to visualize, which are shown then on a Google map. By clicking on each object users can see additional related information, such as the name of the objects and possible associations with external semantic data stores. Objects are passed to the map by using the Google Maps javascript API. The last visualization tool is an Exhibit [34] GUI, an open source-publishing framework for data-rich interactive web pages. Exhibit enables the creation of dynamic exhibits of data collections without resorting to complex databases and server-side technologies. The data collections can be searched and browsed (through advanced text search and filtering functionalities) using faceted browsing.

---

[32] https://www.w3.org/2001/sw/wiki/OpenLink_Virtuoso
[33] http://wit.istc.cnr.it/prisma/webcontent/sparql.php

### 3.3.7 Open Services (City SDK)

This module allows the CitySDK development kit[35] to manage urban reporting. CitySDK is a service development kit for cities and developers that aims at harmonizing application programming interfaces (APIs) across cities. CitySDK APIs enable new services and applications to be rapidly developed, scaled and reused by providing a range of tools and information for both cities and developers. In particular, it leverages the Issue reporting API / Open311 API (and exposes REST web services) to add, edit and remove reports of the City Reporter web application. This module has been developed on top of Talend ESB (for the communication with the City Reporter Backend services in order to get data on urban reporting, which are transformed to CitySDK format) and City Reporter Backend (to directly communicate with the database). Currently this module is needed by the mobile app GeoReporter but in the future we plan to include other features, which will exploit it.

### 3.3.8 Web Applications

This module includes two components: TOUI and City Reporter. TOUI stands for Task Oriented User Interface and provides the interfaces for the management of the process tasks developed on jBPM. City Reporter is a front-end that allows users (citizens) to insert urban reports and visualize the state of all the reports (e.g. whether the report has been processed or not). Thus, TOUI is a user interface for public employers that need to handle urban reports sent by citizens using City Reporter. Both components have been developed using HTML5, AngularJS[36] (one of the most used open source framework to develop web applications), PHP and Bootstrap[37]. All of these technologies are responsive and use REST web services for communication purposes. The TOUI component communicates with CAS for the single sign-on authentication and with the REST service of Talend ESB. City Reporter Front-end also authenticates through the CAS module and it exchanges information with the City Reporter backend through its REST web services. City Reporter is also provided through a dedicated mobile app (GeoReporter module, also shown within the whole architecture depicted in Figure 1).

### 3.3.9 Webportal module

This module enables the instantiation of a web platform in the SaaS layer. Two components form this module: WordPress and User Management. The User Management component aims at providing services related to the users such as:
- self-registration;
- password reset and change;
- insertion of users by the administrator.

---

[34] http://www.simile-widgets.org/exhibit/
[35] http://www.citysdk.eu/
[36] https://angularjs.org/
[37] http://getbootstrap.com/

The web portal is developed as a CMS, which uses the open source platform WordPress[38]. The User Management component is developed in Javascript, HTML 5 and PHP, and it is integrated in WordPress through a plugin. Wordpress is also connected to the CAS module for the single sign-on authentication. Besides, the User Management component uses LDAP as repository of the users' information.

### 3.3.10 OpenTripPlanner module

OpenTripPlanner[39] is an open source platform for multi-modal and multi-agency journey planning that helps finding itineraries combining transit, pedestrian, bike, and car segments. It provides several map-based web interfaces and REST APIs for the use by third-party applications. It relies on open data standards including GTFS[40] (General Transit Feed Specification) for transit and OpenStreetMap[41] for street networks. The OpenTripPlanner module needs: (i) a map of a reference area and (ii) a GTFS file containing information about public transportation (such as the scheduling of buses, bus stops, etc.)

The module is connected to the GTFS services module that provides files in GTFS format. Moreover it is connected to OpenTripPlanner mobile module that uses it to provide the same functionalities to mobile devices.

### 3.3.11 City Welfare modules

City Welfare allows citizens to find voluntary services that are useful to satisfy their needs. It is possible to specify type, area, and the time when a provision is supplied and then the system returns the organizations that can satisfy those requirements. It is also possible for users to browse and search the registries of voluntary organizations. Finally, the system provides a tool to design new voluntary services involving more organizations and stores search services in order to build up a map of user requirements. City Welfare consists of three architectural layers: (i) a front-end developed using HTML5, AngularJS, PHP, and Bootstrap; (ii) a back-end developed in Java, using Spring and Hibernate frameworks; and (iii) a MYSQL database.

This module is directly integrated with CAS. Data regarding users and groups are stored there through the LDAP module.

### 3.4 Prisma Solution (SaaS Layer)

Once the backbone (IaaS and PaaS layers) of the architecture has been described, we discuss here what the SaaS layer consists of and some test cases we have instanced on existing municipalities. Whereas the PaaS layer can be compared to the *class definition* in the object-oriented programming language domain, the SaaS layer can be considered as the *instance of a class* defined in the PaaS layer. In more detail, in the SaaS layer there are the following four main components: Smart City Portal, Reporter, Movers and Welfare. These components represent the instances of the underneath modules and can be employed by any PA or organization to handle its urban and citizens data. Therefore, Smart City Portal is the specific instance of the Web Portal and Opendata modules; Smart City Reporter is the instance of the Web Applications and Mobile App modules; Smart City Movers is the instance of the OpenTripPlanner mobile module, and Smart City Welfare is the instance of the City Welfare web component. The instances of the applications within the SaaS layer can be configured in any languages. Examples of applications we have developed have been setup in Italian (the following figures contain some examples). The Smart City Portal has been instanced on two different PAs, one at the municipality of Catania[42] and another one at the municipality of Syracuse[43]. The same operation has been done for these cities for Smart City Reporter (both the City Reporter[44] and the TOUI components[45]). The Smart City Portal and Smart City Welfare can be publicly accessible (successively URLs are provided). The other websites of the SaaS layer are protected by password since they are still under development; however some screenshots are illustrated in the following. Figure 3 shows a screenshot of the Smart City Reporter used in the municipality of Syracuse where any citizen can insert a new report, specifying the type of report, the place referred in the report, a description, an image, and the localization in a map). Figure 4 shows the list of reports with basic information and their status (a report can be in progress, not opened yet, or closed).

---

[38] https://wordpress.org/
[39] http://www.opentripplanner.org/
[40] https://developers.google.com/transit/gtfs/
[41] http://www.opentripplanner.org/
[42] https://www.catania-smartcity.it/

**Figure 2 – City Reporter Front-end, insert operation of a new report**

---

[43] http://www.siracusa-smartcity.it/
[44] City Reporter for the city of Catania https://www.catania-smartcity.it/cityreporter/, City reporter for the city of Syracuse https://www.siracusa-smartcity.it/cityreporter/
[45] TOUI for the city of Catania https://www.catania-smartcity.it/toui, TOUI for the city of Syracuse https://www.siracusa-smartcity.it/toui

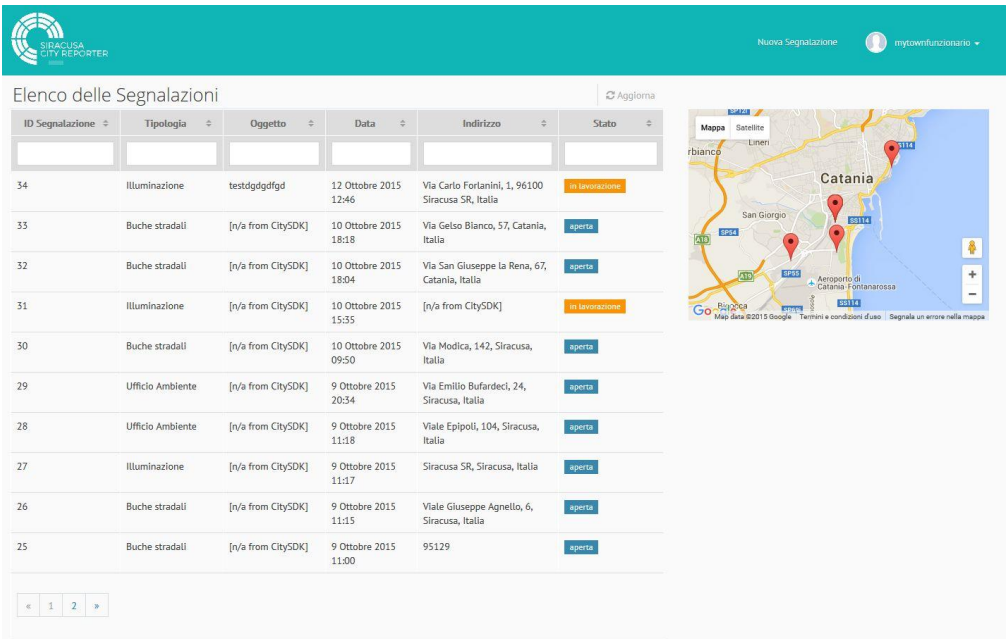

**Figure 3 – List of current reports and their status**

Figure 5 shows a screenshot of the TOUI used for the case of the municipality of Syracuse where, as an example, a list of activities is displayed to a public employee. Figure 7 shows the details of one of the activities.

Two instances of the OpenTripPlanner module, representing the Smart City Movers application, have been allocated for the two municipalities of Catania and Syracuse. They are already available online[46] and ready to be used by any citizens and stakeholders. Figure 6 shows the instance of the Smart City Movers application for the case of Catania.

---

[46] Smart City Movers for Catania: https://www.catania-smartcity.it/citymover/, Smart City Movers for Syracuse: https://www.siracusa-smartcity.it/citymover/

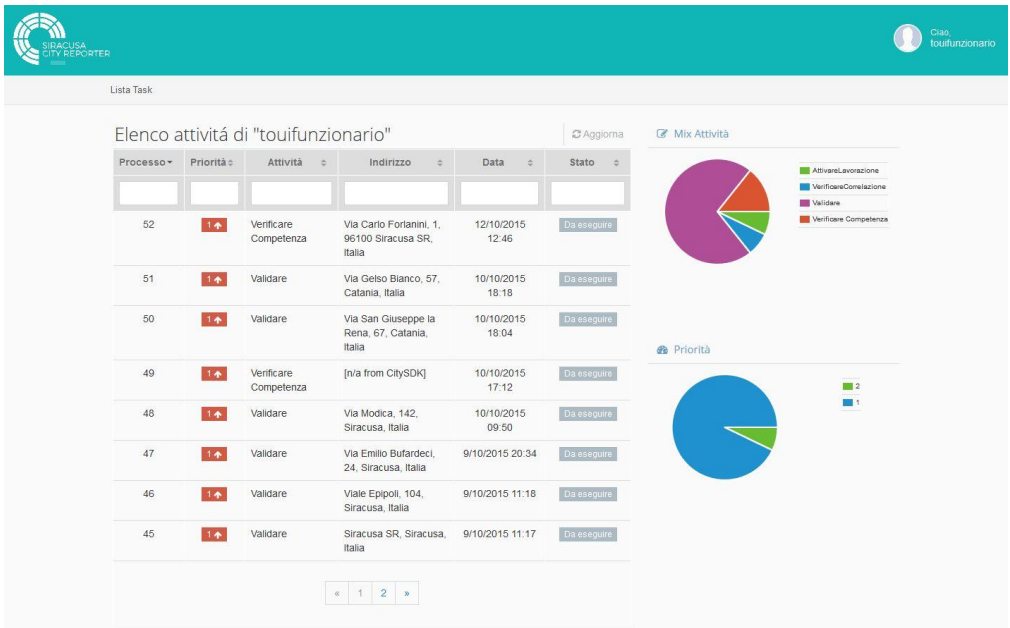

**Figure 4 – TOUI list of activities (id, priority, description, address, date, status)**

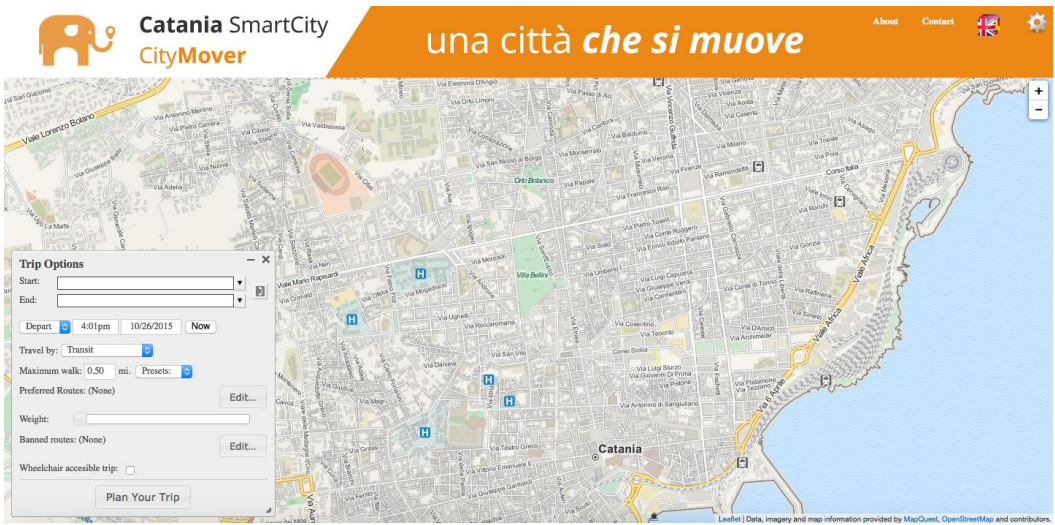

**Figure 5 - Smart City Movers for the municipality of Catania**

**Figure 6 – Details of one activity of Figure 5**

Finally, Figure 8 depicts a screenshot of the mobile app of the City Reporter for the Syracuse use case. The implementation and instantiation of the City Welfare system is publicly available[47]. By using it, a citizen, or in general any enabled web user, can see, search, and/or browse the list of organizations (see Section 3.3.11) and check the related details.

---

[47] https://www.catania-smartcity.it/CityWelfare/.

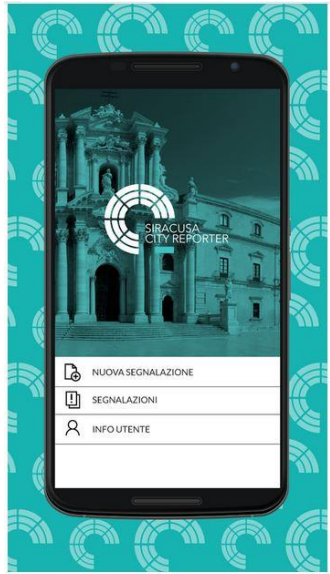

**Figure 7 – The GeoReporter mobile app for the city of Syracuse**

### 3.5 Discussion on the benefits for PAs in using the proposed platform

The size of the current residential construction industry market is characterized by traditional activities that are deeply rooted in the management and production of the majority of SMEs operating in Sicily, with limited ability to create induced products.

Within the context of this work, three lines of actions have characterized the development of the proposed platform. These can significantly contribute to the modernization of the structure and production processes of PAs; they assume the use of technologically advanced solutions as added value in that modernization, and in the direct comparison with market areas increasingly integrated with ICT. Thus, the first line of action relates to the design of the platform that is entirely guided by the open source principles. According to national and European laws, before acquiring new software, PAs are required to conduct a comparative analysis of the different available solutions, taking into account a number of criteria such as costs of the software, use of open interfaces, the level of guaranteed interoperability, etc. Where this analysis highlights the impossibility to adopt an open source solution, PAs can ultimately use commercial software. In the Italian context, where our platform has been deployed, the Agency for Digital Italy (AgID) of the Presidency of the Council of Ministers, in order to implement the earlier mentioned regulations, recommends eliminating any legacy solutions in favor of a model widely distributed and based on open interfaces. The use of open source software, possibly shared among PAs, can also have a positive impact on the costs paid by PAs themselves, and in sustaining small PAs in their innovation programmes: PAs can modify and integrate where necessary the programming code, making it available for the use by others PAs that need to address similar issues. This inevitably can lead to a larger collaboration between public bodies with a potential increase of the quality of PAs' business processes.

The second line of action regards the cloud computing paradigm, the pillar of the entire design and development of the PRISMA platform. Cloud computing can additionally contribute to reduce costs of PAs while maintaining interoperability and portability requirements. It has been widely discussed that PAs can take advantage from the usage of cloud computing infrastructures in order to realize economies of scale. In Europe, the European Commission also tries to measure such advantages with a set of indicators included in the so-called DESI (Digital Economy and Society

Index) framework (Commission, 2015). In Italy, the number of datacenters owned and managed by PAs, often very old, dispersed, and difficult to maintain up-to-date, is not precisely known; from an evaluation provided by AgID [48] it turned out that there exist approximately 4.000 datacenters distributed in the Italian territory, with more than 20.000 installed servers and just as many personalized PAs' software solutions. In this scenario, the need of consolidation and rationalization of ICT hardware and software structures is dramatically rising. Cloud computing can be a crucial paradigm to be used in such a fragmented context, thanks to its inherent advantages deriving from a possible CAPEX (CApital EXpenditure) reduction and OPEX (OPerational EXpenditure) optimization, employed by the pay-as-you-go model. In addition, adopting cloud computing paves the way towards new forms of collaboration among PAs: this is the case of the two real case examples of Catania and Syracuse, but it is also emerging at the different governance levels of PAs, where central or large local administrations are consolidating their datacenters, providing a set of services for the use by smaller PAs that, in recent crisis periods, would have been no longer able to sustain the costs of maintenance of their own old datacenters. PRISMA platform can therefore constitute a successful infrastructure to be widely re-used in such a scenario. Despite the advantages of the cloud computing's economies of scale, there are still some challenges regarding both interoperability of the different implementation of cloud service models (IaaS, PaaS, SaaS) and security. Whereas security in cloud computing is still an ongoing research issue, interoperability has been recently considered. In particular, OpenStack, which is the main implementation of the IaaS layer of the PRISMA platform, is starting being seen as the IaaS standard de facto due to its maturity, diffusion, and openness nature.

Last but not least, the third line of action relates to the ability of the PRISMA platform to make use of open data and open services. With the relatively recent revision of the European Public Sector Information (PSI) Directive 2013/37/EU[49], PAs are required to make available documents and data according to the principles of the open data paradigm (open and machine-readable formats, open licenses, free-of charge or marginal costs business models applied to data, re-use of the data). Freeing up data can have a positive economic impact. The study conducted by McKensey in (Manyika J. et al, 2013) highlights that "*open data have the potential to enable more than $3 trillion in additional value annually*" across a number of domains, thus enabling a substantial economic growth. These results have been recently mentioned in other studies that aim to quantify the economic impact of open data (ODI, 2015). In these studies it is clearly stated that open data may lead organizations to better use existing resources and facilitate the creation of innovative applications and services. The proposed platform, in fact, on the top of the SaaS layer, offers a set of APIs that can be used to interrogate the data, thus clearly stimulating the development of additional services and applications by local developers, stakeholders and citizens. It supports a new model of a digital ecosystem based on the activities of a PA and enables new methods of interactions between different PAs, citizens, industries and other institutions. However, to effectively benefit from the promised economic growth, it is crucial to concentrate on quality aspects of open data and its sustainability in the long-term. In this sense, the use of Semantic Web standards and technologies, such as those employed by the proposed platform, eases the development of a real data culture in the public sector: administrations are more focused on improving the managed data and the processes underneath data production and publication phases rather than on the production of services, which can be left to enterprises of various nature, large as well as small and medium enterprises. Finally, publishing open government data allows citizens to control the actions of PAs and to participate actively in city life and administration, by providing comments or feedbacks that can be useful towards a better organization and management of the public sector.

---

[48] AgID, "Guidelines for the rationalization of the digital infrastructure of the Italian Public Administration",
http://archivio.digitpa.gov.it/sites/default/files//Linee%20guida%20razionalizzazione%20CED%20PA.pdf
(in Italian)

[49] http://eur-lex.europa.eu/legal-content/EN/TXT/HTML/?uri=CELEX:02003L0098-20130717&from=EN

## 4. CONCLUDING REMARKS

In this paper we described an open source, cloud, interoperable ICT platform for smart government developed within the PRISMA project for the municipalities of Catania and Syracuse. The platform includes the three typical layers of cloud-computing architectures (IaaS, PaaS, SaaS) and several components that expose APIs, mostly based on REST and SOAP web services, to promote a wide usage of them by developers, stakeholders and local communities. IaaS, the modules of the PaaS and SaaS have been described in detail, showing the adopted solution and the rationale behind the use of open source technology. Three principal goals of the platform are (i) to simplify complex procedures employed within PAs, (ii) to encourage local developers and businessmen to develop services using APIs of the platform giving a strong impact to local economies of PAs, (iii) to stimulate citizens to be more actively involved within city life by providing feedbacks, suggestions or reports for a wide spectrum of activities.

Two Sicilian municipalities, Catania and Syracuse, are already using the presented platform and several others in Italy expressed their interest in doing the same. Open source software has been used in order to ease the adoption of the platform by any other municipality. Discussions on the potential economic impact of the platform have been included, focusing on three different lines of actions that have characterized the development of the platform. Although feedback and statistics we received for the presented urban framework are still preliminary, the trend we have noticed so far makes us optimistic.

PRISMA is going to end at the beginning of 2016 and one of the goals of the project was to create a community that will maintain the platform and its source code available within a web site that will be created and that will include a forum, a wiki and all the best practices of the open source paradigm.

Future work includes the creation of an Urban App Store that can provide applications developed on the top of the proposed platform with the aim of constructing and shaping an Urban Smart Community offering additional benefits to local PAs.

## ACKNOWLEDGMENTS

This work has been supported by the PON R&C project PRISMA, "PlatfoRms Interoperable cloud for SMArt-Government", ref. PON04a2 A Smart Cities, under the National Operational Programme for Research and Competitiveness 2007-2013. Moreover, this work has been supported by the AW4City2015 workshop[50] held in Florence, Italy within the WWW 2015 conference.

---

[50] https://aw4city.wordpress.com/